\documentclass[aps,showpacsx,twocolumn]{revtex4}
\usepackage{graphicx}
\usepackage{bm}
\usepackage{color}
\usepackage{eufrak}
\usepackage{mathrsfs}

\begin{document}

\title{Magnetic Field Effect on Charmonium Production in High Energy Nuclear Collisions}
\author{Xingyu Guo$^1$, Shuzhe Shi$^1$, Nu Xu$^{2,3}$, Zhe Xu$^1$, Pengfei Zhuang$^1$}
\affiliation{$^1$Physics Department, Tsinghua University and Collaborative Innovation Center of Quantum Matter, Beijing 100084, China\\
         $^2$Nuclear Science Division, Lawrence Berkeley National Laboratory, Berkeley, CA 94720, USA\\
         $^3$Key Laboratory of Quark and Lepton Physics (MOE) and Institute of Particle Physics, Central China Normal University, Wuhan 430079, China}
\date{\today}
\begin{abstract}
It is important to understand the strong external magnetic field generated at the very beginning of high energy nuclear collisions. We study the effect of the magnetic field on the charmonium yield and anisotropic distribution in Pb+Pb collisions at the LHC energy. The time dependent Schr\"odinger equation is employed to describe the motion of $c\bar c$ pairs. We compare our model prediction of non-collective anisotropic parameter $v_2$ of $J/\psi$s with CMS data at high transverse momentum. This is the first attempt to measure the magnetic field in high energy nuclear collisions.
\end{abstract}
\maketitle
%%%%%%%%%%%%%%%%%%
It is widely accepted that the most strong magnetic field in nature can be generated in the early stage of relativistic heavy ion collisions. The peaked magnitude of the field can reach $eB \sim 5 m_\pi^2$ for semi-central Au+Au collisions at the Relativistic Heavy Ion Collider (RHIC) and $70 m_\pi^2$ for semi-central Pb+Pb collisions at the Large Hadron Collider (LHC)~\cite{rafelski,skokov,asakawa,voronyuk,ou,deng,liu}, where $e$ is the electron charge and $m_\pi$ the pion mass. Considering the interaction between the field and the hot medium formed in the later stage of the collisions, one could study fundamental QCD topological structures~\cite{gursoy} in hot and dense nuclear matter such as chiral magnetic effect~\cite{vilenkin,metlitshi,newman,kharzeev1}, chiral magnetic wave~\cite{kharzeev2,gorbar,burnier}, chiral separation effect~\cite{kharzeev3,kharzeev4,fukushima}, chiral vortical effect~\cite{kharzeev5,kharzeev6}, chiral electric separation effect~\cite{huang}, and enhancement of elliptic flow of charged particles~\cite{tuchin,basar1,basar2}. Recent experimental results on the study of the QCD intrinsic properties at RHIC and LHC can be found in Ref.~\cite{xuzhaoshou}.

Although we have crude estimation on the strength of the initial magnetic field, it may decay very fast and survive only in the very beginning of heavy ion collisions. So far, no experimental determination on the magnetic field has been made. In this Letter, we propose using the anisotropic production of high momentum charmonia to probe the initial magnetic field in high energy nuclear collisions. Due to their large masses, heavy quarks are produced in the early stage through hard scatterings. For low momentum charmonia, they can be produced in the initial stage and regenerated~\cite{pbm,rapp,thews,yan} in the hot medium, and both will be suppressed through Debye screening~\cite{matsui}. High momentum charmonia, on the other hand, are purely formed before the formation of the hot medium. Except interaction with the magnetic field, which will be discussed here, the high momentum charmonia are almost not affected by the hot medium and therefore can be used as an ideal tool to probe the initial magnetic field in high energy nuclear collisions. Note, in this work we only focus on the spectators induced filed. The effects of the delayed decay of the field~\cite{gursoy} caused by the conducting medium will not change our main conclusions.

Since the life time of the magnetic field $t_B \sim 0.1$ fm/c~\cite{deng} is much shorter than the charmorium formation time $t_f \sim 0.5$ fm/c~\cite{gerschel} from a $c\bar c$ pair to a state $\Psi=J/\psi,\psi',\chi_c,\cdots$, the interaction between the $c\bar c$ pair and the field could cause some profound effects on (i) Charmonium yields:  the magnetic field induced force will change the charmonium fractions $|C_\Psi|^2$ in the $c\bar c$ pair $|c\bar c\rangle =C_\Psi|\Psi\rangle$ and thus alters the relative yields among different charmonium states, and (ii) Charmonium distribution: the specific direction along the magnetic field breaks down the rotational symmetry of the system, which leads to an anisotropic charmonium production in the transverse plane. Both effects can be experimentally checked.

As charm quarks are heavy enough in comparison with the inner movement inside a charmonium bound state which is determined by the $J/\psi$ radius due to the uncertainty principle, $m_c\sim 1.3$ GeV $>p\sim 1/r=1/(0.5\text {fm})=0.4$ GeV, we can ignore the relativistic effect in considering the inner structure of a charmonium. We employ the time dependent Schr\"odinger equation to describe the evolution of a $c\bar c$ pair wave function $\Phi(t,{\bf r}_c,{\bf r}_{\bar c})$,
\begin{equation}
\label{schroedinger}
i{\partial\over \partial t}\Phi=\hat H\Phi
\end{equation}
with the Hamiltonian operator
\begin{equation}
\label{hamiltonia}
\hat H={\left(\hat {\bf p}_c-q{\bf A}_c\right)^2\over 2m_c}+ {\left(\hat {\bf p}_{\bar c}+q{\bf A}_{\bar c}\right)^2\over 2m_c}+V,
\end{equation}
where ${\bf A}$ is the magnetic potential, $q=2e/3$ the charm quark electron charge, and $V$ the potential between the quark and antiquark. We take the Cornell potential together with the spin-spin interaction~\cite{lattice},
\begin{equation}
V(r) = -\frac{\alpha}{r} + \sigma r + \beta e^{-\gamma r} {\bf s}_c\cdot{\bf s}_{\bar c}.
\end{equation}
By fitting the charmonium spectrum in vacuum, one can fix the potential parameters as $m_c=1.29$ GeV, $\sigma=0.174$ GeV$^2$, $\alpha=0.312$, $\beta=1.982$ GeV and $\gamma=2.06$ GeV.

By separating the $c\bar c$ wave function into a center-of-mass part and a relative part $\Phi=\Phi_R\Phi_r$ with the center-of-mass coordinator ${\bf R}=({\bf r}_c+{\bf r}_{\bar c})/2$ and relative coordinator ${\bf r}={\bf r}_c-{\bf r}_{\bar c}$, and further expanding the relative part in terms of the charmonium wave functions $\Psi({\bf r})$,
\begin{equation}
\label{expansion}
\Phi = {1\over \sqrt{2\pi}}e^{i{\bf P}_k\cdot{\bf R}-i{{\bf P}_p^2\over 4m_c}t}\sum_\Psi C_\Psi e^{-iE_\Psi t} \Psi,
\end{equation}
the probability $|C_\psi(t)|^2$ for the $c\bar c$ pair to be in the charmonium state $\Psi$ satisfies the evolution equation
\begin{equation}
\label{cpsi}
{d\over dt}C_\Psi=\sum_{\Psi'}e^{i(E_\psi-E_{\psi'})t}C_{\Psi'}\int d^3{\bf r}\Psi^*({\bf r})\hat H_B\Psi'({\bf r}),
\end{equation}
where we have separated the Hamiltonian into a vacuum part and a magnetic field dependent part,
\begin{equation}
\label{h}
\hat H=\hat H_0 + \hat H_B,
\end{equation}
with
\begin{equation}
\label{hb}
\hat H_B= {-\boldsymbol \mu}\cdot {\bf B}-{q\over 2m_c}(\hat {\bf P}_p\times {\bf B})\cdot{\bf r}+{q^2\over 4m_c}({\bf B}\times {\bf r})^2.
\end{equation}
The charmonium energy $E_\psi$ and wave function $\Psi({\bf r})$ in Eqs.(\ref{expansion}) and (\ref{cpsi}) are determined by $\hat H_0$,
\begin{equation}
\label{h0}
\hat H_0\Psi=E_\Psi \Psi.
\end{equation}

Since we have now a special direction, the direction of ${\bf B}$, the kinetic momentum ${\bf P}_k={\bf P}-q({\bf A}_c-{\bf A}_{\bar c})$ with the total momentum ${\bf P}={\bf p}_c+{\bf p}_{\bar c}$ is no longer conserved, and the conserved momentum is the pseudo momentum ${\bf P}_p = {\bf P}+q({\bf A}_c-{\bf A}_{\bar c})$~\cite{alford}. In Eq.(\ref{hb}) the $c\bar c$ pair interaction with the magnetic field is manifested by three terms: the first term is the spin-field interaction with the spin magnetic moment ${\boldsymbol \mu}=q/m_c({\bf s}_c-{\bf s}_{\bar c})$, the second term comes from the Lorentz force which is proportional to the $c\bar c$ momentum, and the third term is the harmonic potential which is quadratic in $q{\bf B}$ and therefore its effect is much smaller in comparison with the first and second terms which are linear in $q{\bf B}$. At high momentum, the Lorentz force is the dominant magnetic field effect. Note that the spin-field interaction makes the spin angular momentum no longer a conserved quantity, the spin singlet $\eta_c$ will couple with one of the triplet $J/\psi$ which carries zero spin component along the magnetic field $s_B=0$~\cite{alford,yang}. On the other hand, the Lorentz force and the harmonic potential breaks the rotational symmetry in the coordinate space.

Let us take the nuclear colliding direction as the $z$-axis and the impact parameter ${\bf b}$ parallel to the $x$-axis. While the created magnetic field in heavy ion collisions depends on the events, and the magnitude and direction of the field fluctuate in space and time~\cite{rafelski,skokov,asakawa,voronyuk,ou,deng,liu}, we consider here an averaged magnetic field ${\bf B}$ along the $y-$axis in the space-time region determined by the colliding energy and nuclear geometry,
\begin{equation}
\label{b}
{\bf B} = \left\{ \begin{array}{ll}
B{\bf e}_y, & 0<t<t_B, \frac{x^2}{(R_A-b/2)^2}+\frac{y^2}{(b/2)^2}+\frac{\gamma_c^2z^2}{(b/2)^2} < 1, \\
0, & \text {others}. \\
\end{array}\right.
\end{equation}
which, from the relation ${\bf B}=\nabla \times {\bf A}$, leads to $A_x=Bz$ and $A_y=A_z=0$. For Pb+Pb collisions with centrality $40\%$ and at LHC energy $\sqrt{s_{NN}}=2.76$ TeV, the geometry parameters and the Lorentz factor are fixed to be $R_A=6.6$ fm, $b=8$ fm  and $\gamma_c=1400$. From the MC simulation~\cite{deng}, we take the magnitude and the life time of the magnetic field $eB=25 m_\pi^2$ and $t_B=0.2$ fm/c. At RHIC energy, the initially created magnetic field is much weaker, $eB\simeq m_\pi^2$~\cite{deng}. If the field decays very fast before the formation of the hot medium where the Faraday induction may prolong the field~\cite{gursoy}, the magnetic field effect on the charmonium production can be safely neglected. In the following we will focus on the magnetic field effect at LHC energy.

To solve the dynamical equation (\ref{cpsi}) for the probability magnitude $C_\Psi(t)$, we need the initial value $C_\Psi(0)$ or the initial wave function $\Phi_r(0)$. Suppose the relative motion of the $c\bar c$ pair is initially described by a compact Gaussian wave package
\begin{equation}
\label{initial}
\Phi_r(0)\sim e^{-{({\bf r}-{\bf r}_0)^2\over \sigma_0^2}}
\end{equation}
with averaged relative coordinate ${\bf r}_0$ and distribution width $\sigma_0$. Since there is no reason to choose a special direction before the magnetic field is introduced, the azimuth angles of ${\bf r}_0=r_0(\sin\theta_0\cos\phi_0,\sin\theta_0\sin\phi_0,\cos\theta_0)$ are randomly distributed, and we do ensemble average over $\theta_0$ and $\phi_0$ in the final state calculations. The remaining two parameters $r_0$ and $\sigma_0$ can be determined by fitting the charmonium fractions in p+p collisions. Suppose an initial point-like wave function $\delta({\bf r}-{\bf r}')$ develops as $e^{({\bf r}-{\bf r}')^2/(v^2t^2)}$ with a constant expansion velocity $v$, the initial Gaussian wave package evolves as
\begin{equation}
\label{phit}
\Phi_r(t) \sim \int d^3{\bf r}' e^{-{({\bf r}'-{\bf r}_0)^2\over \sigma_0^2}}e^{-{({\bf r}-{\bf r}')^2\over v^2t^2}} \sim  e^{-{({\bf r}-{\bf r}_0)^2\over \sigma_t^2}}
\end{equation}
in p+p collisions. It is always a Gaussian wave package but with a time dependent width $\sigma_t^2=\sigma_0^2+v^2t^2$. Calculating the probability $|C_\Psi(t_f)|^2 = |\langle\Psi|\Phi_r(t_f)\rangle|^2$ in vacuum with $B=0$ at the charmonium formation time $t_f$, and taking the experimentally observed decay branch ratios ${\cal B}(\Psi\to J/\psi)$~\cite{pdg}, we obtain the fractions of the direct $J/\psi$ production and feed down from the excited states in the finally observed prompt $J/\psi$s,
\begin{equation}
\label{ratio}
R_\Psi(t)={|C_\Psi(t)|^2{\cal B}(\Psi\to J/\psi)\over \sum_\Psi |C_\Psi(t)|^2{\cal B}(\Psi\to J/\psi)}
\end{equation}
with the definition of ${\cal B}(J/\psi\to J/\psi)=1$. Taking $R(J/\psi)=60\%$, $R(\psi')=10\%$ and $R(\chi_c)=30\%$ from the recent p+p data at LHC energy~\cite{atlas,cms1,lhcb} and the charmonium formation time $t_f=0.5$ fm/c~\cite{gerschel}, the remaining two parameters in the initial wave package $\Phi_r(0)$ are fixed to be $r_0=0.68$ fm and $\sigma_0=0.02$ fm, which correspond to the expansion velocity $v=0.72$c and the width $\sigma_{t_f}=0.38$ fm.

With the known initial wave function $\Phi_r(0)$ or the initial probability magnitudes $C_{J/\psi}(0):C_{\psi'}(0):C_{\chi_c}(0)=1:0.5:1.2$, we can solve the evolution equation (\ref{cpsi}) for $C_\Psi(t)$ in the time region $0<t<t_B$ when the magnetic field exists. After that the wave function evolves according to the expansion rule (\ref{phit}) with $\Phi_r(t_B,{\bf r})$ as the initial condition,
\begin{equation}
\label{phit2}
\Phi_r(t,{\bf r}) \sim \int d^3{\bf r}' \Phi_{r'}(t_B,{\bf r}')e^{-{({\bf r}-{\bf r}')^2\over v^2(t-t_B)^2}}\ \ \ \ \ \ \text {for}\ \ t>t_B.
\end{equation}

We now show our numerical results about the magnetic field effect on the charmonium production in heavy ion collisions. We focus on the central rapidity region in Pb+Pb collisions with impact parameter $b=8$ fm and at LHC energy $\sqrt {s_{NN}}=2.76$ TeV. In our calculation, the initial $c\bar c$ pairs which are determined by the colliding energy and the nuclear geometry are assumed to be distributed homogeneously in the overlapped region of the two colliding nuclei. To have a good precision in solving the coupled equations (\ref{cpsi}) for $C_\Psi(t)$, we take a cutoff in the sum over the charmonium eigenstate $\Psi$: the main quantum number $n\leq 6$ and the orbital angular momentum number $l\leq 7$.

\begin{figure}[!hbt]\centering
\includegraphics[width=0.48\textwidth]{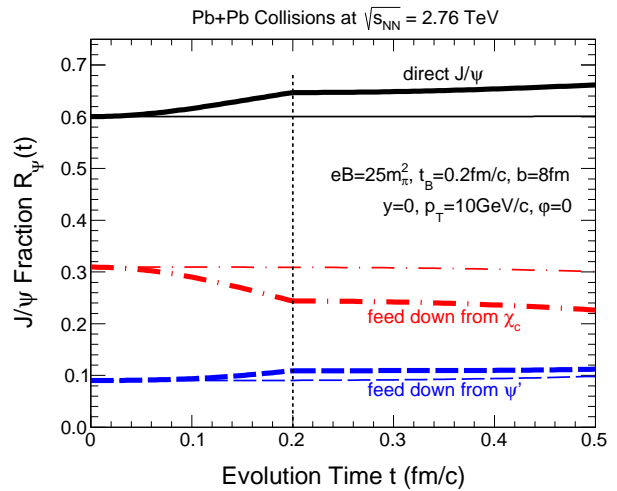}
\caption{(Color online) The time evolution of $J/\psi$s from direct production (solid lines) and feed down from $\psi'$ (dashed lines) and $\chi_c$ (dot-dashed lines). The thick and thin lines represent the results from the calculations with and without the external magnetic filed, respectively.  As indicated by the vertical short-dashed line, the magnetic field only lasts during the time $t<t_B= 0.2$ fm/c. }
\label{fig1}
\end{figure}

In Fig.\ref{fig1}, the $J/\psi$ fractions $R_\Psi(t)$ from different channels are shown as functions of $c\bar c$ evolution time. The results with and without the external magnetic field are displayed by thick and thin lines, respectively. These fractions are extracted at fixed transverse momentum $p_T=10$ GeV/c from Pb+Pb collisions at impact parameter $b=8$ fm and LHC energy $\sqrt s_{NN}=2.76$ TeV. In case there is no magnetic field, the direct $J/\psi$ production and feed down from $\psi'$ are slightly enhanced, while the contribution from $\chi_c$ is weakly suppressed, indicating a $\chi_c$ decay to $J/\psi$ and $\psi'$ with quantum selection rule $\delta l=1$. When the magnetic field is turned on, the converting from the high lying state $\chi_c$ to $J/\psi$ and $\psi'$ becomes much more dramatic, as one can see the thick lines in Fig.\ref{fig1}. Since the magnetic field introduces a special direction, the $c\bar c$ motion  becomes anisotropic. Here we fix the $c\bar c$ azimuthal angle in the transverse plane $\varphi=\arctan (p_y/p_x)=0$ where the magnetic field effect is expected to be the strongest. In the time period $t<t_B$, the external magnetic field serves as a stimulator that enhances the quantum mechanics allowed transition from $\chi_c$ to $J/\psi$ and $\psi'$. After the field is off at $t>t_B$, the variation of all the fractions with time becomes mild again. At the formation time $t_f=0.5$ fm/c, the relative enhancement for both direct $J/\psi$ production and feed down from $\psi'$ are found to be $10\%$, and the contribution from $\chi_c$ decay is relatively suppressed by $23\%$.

\begin{figure}[!hbt]\centering
\includegraphics[width=0.48\textwidth]{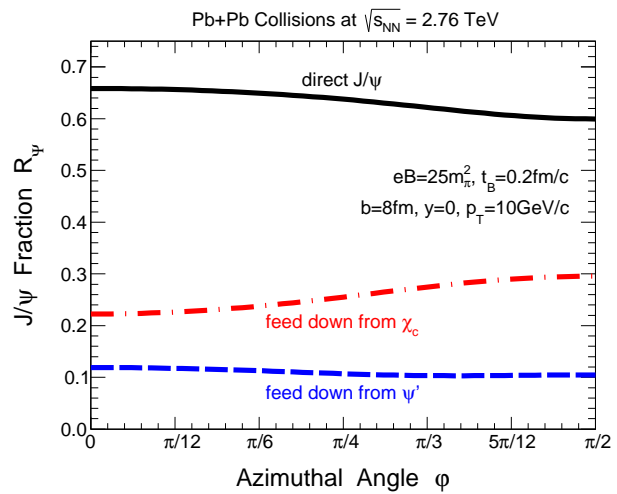}
\caption{(Color online) The magnetic field induced anisotropic behavior of the fractions $R_\Psi$ in the $J/\psi$ yield at the formation time $t_f$. }
\label{fig2}
\end{figure}

Having discussed the different contributions to $J/\psi$ production, we now look at the yields for $J/\psi$, $\psi'$ and $\chi_c$. For $J/\psi$, the ratio between the yields $N_{J/\psi}^{(B)}$ and $N_{J/\psi}^{(0)}$ with and without the magnetic field can be expressed in terms of the probabilities $|C_\Psi^{(B)}(t_f)|^2$ and $|C_\Psi^{(0)}(t_f)|^2$ and the fractions $R_\Psi^{(0)}(t_f)$ at the formation time $t_f$,
\begin{eqnarray}
\label{ratio2}
{N_{J/\psi}^{(B)}\over N_{J/\psi}^{(0)}}&=&{\sum_\Psi |C_\Psi^{(B)}(t_f)|^2{\cal B}(\Psi\to J/\psi)\over \sum_\Psi |C_\Psi^{(0)}(t_f)|^2{\cal B}(\Psi\to J/\psi)}\nonumber\\
&=&\sum_\Psi {|C_\Psi^{(B)}(t_f)|^2\over |C_\Psi^{(0)}(t_f)|^2}R_\Psi^{(0)}(t_f).
\end{eqnarray}
Using the above calculated probabilities and fractions at LHC energy, the maximum $J/\psi$ enhancement at $\varphi=0$ and $p_T=10$ GeV/c is $13\%$. For the excited states, neglecting the feed down from the higher eigen states of $\hat H_0$, we have the yield ratios $N_{\psi'}^{(B)}/N_{\psi'}^{(0)}=|C_{\psi'}^{(B)}(t_f)|^2/|C_{\psi'}^{(0)}(t_f)|^2=1.29$ for $\psi'$ and $N_{\chi_c}^{(B)}/N_{\chi_c}^{(0)}=|C_{\chi_c}^{(B)}(t_f)|^2/|C_{\chi_c}^{(0)}(t_f)|^2=0.84$ for $\chi_c$, which mean, due to the magnetic field, a relative enhancement of $29\%$ and suppression of $16\%$ for $\psi'$ and $\chi_c$ productions in Pb+Pb collisions at the LHC energy.

Fig.\ref{fig2} shows the magnetic field induced anisotropic property of the fractions $R_\Psi(t_f)$ in the transverse plane at charmonium formation time $t_f$. The strength of the Lorentz force acting on the charmonia is most strong at $\varphi=0$, and decreases monotonously with increasing azimuthal angle $\varphi$. Finally at $\varphi=\pi/2$, the force disappears and only the weak harmonic potential exists, the fractions approach to their vacuum values.

\begin{figure}[!hbt]\centering
\includegraphics[width=0.48\textwidth]{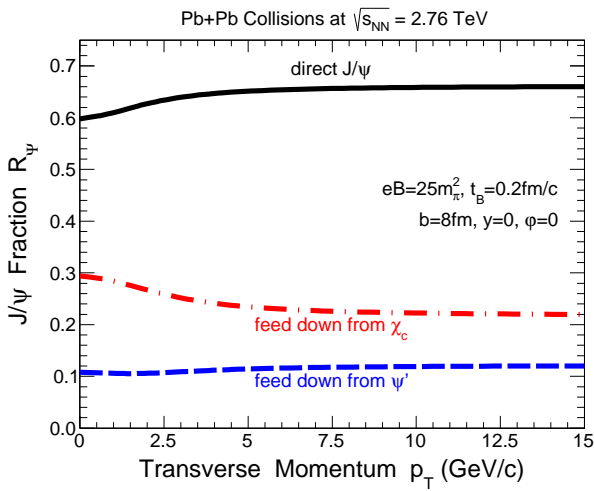}
\caption{(Color online) The magnetic field induced transverse momentum dependence of the fractions $R_\Psi$ in the $J/\psi$ yield at the formation time $t_f$. }
\label{fig3}
\end{figure}

\begin{figure}[!hbt]\centering
\includegraphics[width=0.50\textwidth]{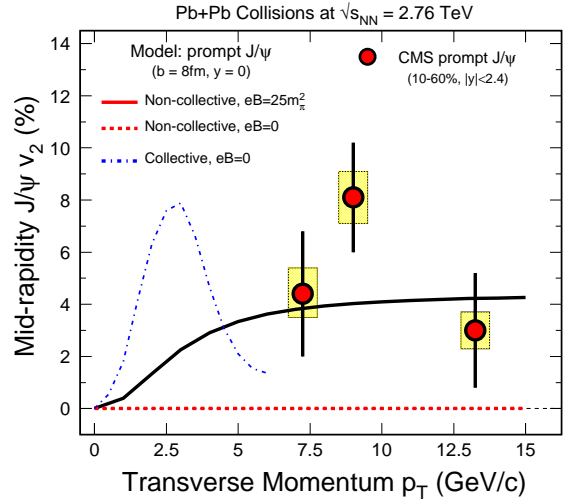}
\caption{(Color online) The transverse momentum dependence of $J/\psi$ $v_2$. The solid and dashed lines are the initially produced non-collective $J/\psi$ $v_2$ with and without magnetic field, and the data at high $p_t$ are from the CMS collaboration~\cite{cms2}. As a comparison, the collective $J/\psi$ $v_2$ from a transport model~\cite{zhou} is shown as the dot-dashed line.}
\label{fig4}
\end{figure}

We now show in Fig.\ref{fig3} the magnetic field induced transverse momentum dependence of the fractions $R_\Psi(t_f)$ at charmonium formation time $t_f$. We again fix the angle $\varphi=0$ to see the maximum magnetic field effect. At $p_T=0$ the Lorentz force disappears, the slight deviation of the fractions from the vacuum values comes from the weak harmonic potential. The strength of the Lorentz force increases with $p_T$, and the change of the fractions for all channels becomes larger till $p_T \sim 3$ GeV/c. In higher $p_T$ region, the change stops and the transverse momentum dependence seems flat. Although the force is increasing with $p_T$, the higher the $p_T$ the shorter the time the $c\bar c$ pair experiences the field. The combination of the momentum dependence of the Lorentz force and the finite time of the external magnetic field leads to the observed saturation in Fig.\ref{fig3}.

Usually, the observed event anisotropy $v_2$ is discussed in the framework of hydrodynamics, representing the collective motion of the medium created in high energy nuclear collisions~\cite{zhou,rapp2,gossiaux}. At the LHC energy, $J/\psi$ $v_2$ has been reported in the region of $p_T<10$ GeV/c~\cite{alice}. Since the collective motion is dominated by the bulk interactions in relatively low momentum region, those high $p_T$ charmonia generated in the initial stage are not expected to be sensitive to the nature of the hot medium. However, the Lorentz force induced anisotropic production in the transverse plane, shown in Fig.\ref{fig2}, may result in a non-collective $J/\psi$ $v_2$ at high $p_T$. The $J/\psi$ $v_2$ as a function of rapidity and transverse momentum is defined as
\begin{equation}
\label{v2}
v_2(\eta,p_t) = \frac{\int_0^{2\pi} N_{J/\psi}(\eta,p_t,\varphi)\cos(2\varphi) d\varphi}{\int_0^{2\pi} N_{J/\psi}(\eta,p_t,\varphi) d\varphi}.
\end{equation}
The numerical result at central rapidity is shown in Fig.\ref{fig4}. In normal calculations without considering the magnetic field~\cite{zhou,rapp2,gossiaux}, the initial $J/\psi$s are isotropically produced with $v_2=0$ (dashed line). However, due to the magnetic field induced Lorentz force, the high $p_T$ $J/\psi$s acquire sizable non-collective $v_2$ (solid line), which explains reasonably well the CMS data~\cite{cms2} for prompt $J/\psi$s in the high $p_T$ region. As a comparison, We show in Fig.\ref{fig4} also the collective $v_2$ (dot-dashed line) calculated from a dynamical transport model~\cite{zhou}. Different from the collective flow which comes from the $J/\psi$ regeneration at low and intermediate $p_T$, the non-collective $v_2$ induced by the magnetic field is mainly in high $p_T$ region. As one can see in Figs.\ref{fig1}, \ref{fig2} and \ref{fig3}, $\chi_c$ is suppressed, corresponding to the $J/\psi$ and $\psi'$ enhancement. Hence it is interesting to point out that if the observed non-collective $v_2$ for $J/\psi$ is positive, the high $p_T$ $v_2$ for $\chi_c$ should be negative.

In summary, by solving the time dependent Schr\"odinger equation for the $c\bar c$ pairs, we discussed the effects of the initially created magnetic field on the relative yields and anisotropic distributions of the charmonium states in high energy nuclear collisions. For Pb+Pb collisions at LHC, we found (i) the directly produced and $\psi'$ decayed $J/\psi$s are enhanced, while the $J/\psi$s from $\chi_c$ decay are suppressed, and (ii) the non-collective $J/\psi$ $v_2$ at large transverse momentum, $p_T \geq 8 $ GeV/c, is as large as $4\%$ and comparable with the CMS data. We wish to point out that in high energy nuclear collisions the initial magnetic field is related to several important measurements reflecting the intrinsic structure of the QCD. The measurement of the strong $J/\psi$ anisotropy at large $p_T$ can be used to diagnose the magnetic filed in such collisions.

\noindent {\bf Acknowledgement:} The work is supported by the NSFC, MOST and DOE grant Nos. 11275103, 11335005, 2013CB922000, 2014CB845400, 2015CB856900 and DE-AC03-76SF00098.

\end{document}